\documentclass[a4paper,fleqn,usenatbib]{mnras}
\usepackage{newtxtext,newtxmath}
\usepackage[T1]{fontenc}
\usepackage{ae,aecompl}
\usepackage{graphicx}	
\usepackage{amsmath}	

\usepackage{amssymb}	
\usepackage{mathrsfs}
\usepackage{longtable}
\usepackage{hyperref}
\usepackage[font=small]{caption}
\usepackage{bm}

\title[AT2018cow]{An environmental analysis of the fast transient AT2018cow and implications for its progenitor and late-time brightness}

\author[N.-C. Sun et al.]{Ning-Chen Sun$^{1, 2, 3}$\thanks{E-mail: sunnc@ucas.ac.cn}, Justyn R. Maund$^3$, Yali Shao$^4$ and Ida A. Janiak$^{3, 5}$ \\
1 School of Astronomy and Space Science, University of Chinese Academy of Sciences, 19A Yuquan Road, Shijingshan District, Beijing 100049, China \\
2 National Astronomical Observatories, Chinese Academy of Sciences, 20A Datun Road, Chaoyang District, Beijing 100101, China \\
3 Department of Physics and Astronomy, University of Sheffield, Hicks Building, Hounsfield Road, Sheffield S3 7RH, UK \\
4 Max-Planck-Institut f\"ur Radioastronomie, Auf dem H\"{u}gel 69, 53121 Bonn, Germany \\
5 Department of Physics and Astronomy, the University of Manchester, Schuster Building, Oxford Road, Manchester, M13 9PL, United Kingdom}
\date{Accepted XXX. Received YYY; in original form ZZZ}

\pubyear{2023}

\begin{document}
\label{firstpage}
\pagerange{\pageref{firstpage}--\pageref{lastpage}}
\maketitle

\begin{abstract}

The nature of the newly discovered fast blue optical transients (FBOTs) is still puzzling astronomers. In this paper we carry out a comprehensive analysis of the molecular gas, ionized gas and stellar populations in the environment of the nearby FBOT AT2018cow based on ALMA, VLT/MUSE and HST/WFC3 observations. A prominent molecular concentration of 6~($\pm$ 1) $\times$ 10$^6$~$M_\odot$ is found in the vicinity of AT2018cow, which has given rise to two active star-forming complexes with ages of 4 $\pm$ 1~Myr and $\lesssim$2.5~Myr, respectively. Each star-forming complex has a stellar mass of 3 $\times$ 10$^5$~$M_\odot$ and has photoionized a giant H~\textsc{ii} region with H$\alpha$ luminosity even comparable to that of the 30~Dor mini-starburst region. AT2018cow is spatially coincident with one of the star-forming complexes; however, it is most likely to reside in its foreground since it has a much smaller extinction than the complex. Its progenitor could have been formed in a different star-forming event, and the non-detection of the associated stellar population constrains the progenitor's age to be $\gtrsim$10~Myr and initial mass to be $\lesssim$ 20~$M_\odot$. We further find the late-time brightness of AT2018cow is unlikely to be a stellar object. Its brightness has slightly declined from 2~yr to 4~yr after explosion and is most likely to originate from AT2018cow itself due to some powering mechanism still working at such late times.

\end{abstract}


\section{Introduction}
\label{intro.sec}

Fast blue optical transients (FBOTs) are among the most important discoveries made by modern time-domain surveys at high cadence. Their peak luminosities are similar to those of normal or superluminous supernovae (SNe); however, they rise to peak in just a few days and then drop rapidly, evolving at a rate much faster than typical SNe \citep{Ho2021}. Pure radioactive decay, which powers most of the normal SNe \citep{Arnett1982}, is unable to explain their fast-evolving light curves \citep{Xiang2021}.They have very blue colours consistent with high photospheric temperatures of $>$10$^4$~K \citep{Kuin2019, Perley2019, Ho2020, Xiang2021}, and many of them are found to be luminous at radio and X-ray wavelengths \citep{Ho2019, Margutti2019, Coppejans2020, Bright2022, Yao2022}.  All these features suggest that FBOTs are a distinct new type of transients that was unknown to us.

It is still unclear what progenitors give rise to FBOTs and what processes power their rapid evolution. Most of the current models can be classified into two categories. The first set of models consider a star being tidally disrupted and accreted onto a black hole \citep[e.g.][]{Liu2018, Kuin2019, Perley2019, Metzger2022}, while the second category regards FBOTs as arising from the core collapse of massive stars. To power the rapid evolution, the collapsar models require some special energy input by, e.g., magnetar \citep{Prentice2018, Fang2019, Liu2022}, jets \citep{Gottlieb2022, Soker2022}, or the strong interaction between ejecta and circumstellar material (CSM; \citealt{Fox2019, Leung2020, Xiang2021}). Some other models consider the accretion-induced collapse of white dwarfs \citep{Wang2020}, the cooling emission from an extended envelope \citep{Drout2014}, or the shock breakout from a dense wind \citep{Rest2018}. Many observational studies \citep{Ho2021, Xiang2021} reveal the similarities in the light curves and spectra between FBOTs and the Type~Ibn SNe (i.e. the explosion of hydrogen-poor stars, possibly stripped in binaries, with strong ejecta-CSM interaction; \citealt{Pastorello2007, Pastorello2008, Hosseinzadeh2017, Sun2020a}), and theoretically, \citet{Dessart2022} showed that the standard-energy explosions of helium stars within dense CSM are reminiscent of FBOTs while the low-energy explosions are more consistent with Type~Ibn SNe. Some other astronomers suggest that FBOTs have mildly or sub-relativistic outflows \citep[e.g.][]{Coppejans2020, Ho2020}, which could bridge the gap between normal SNe and the long gamma-ray bursts with very powerful jets.

At a distance of 63~Mpc, AT2018cow is the closest FBOT ever detected and was monitored intensively soon after its discovery. It has a high peak luminosity of 10$^{44}$~erg~s$^{-1}$ and was bright not only at optical but also at radio, infrared, ultraviolet (UV) and X-ray wavelengths \citep{Kuin2019, Fox2019, Ho2019, Margutti2019, Perley2019, Xiang2021}. The non-detection of linear polarization at 230~GHz possibly suggests a dense circumstellar environment with a strong magnetic field that leads to Faraday depolarization and may support a stellar explosion scenario for AT2018cow \citep{Huang2019}. With very-long-baseline interferometry, \citet{Mohan2020} did not find any evidence for jetted ejecta and argued that AT2018cow could be powered by a magnetar. \citet{Pasham2021} reported a quasi-periodic oscillation in its soft X-ray emission, suggesting the possible presence of a compact object. More recently, \citet{Sun2022b} discovered a hot and luminous source at the position of AT2018cow on images taken at 2--3~yrs after explosion by the Wide Field Camera~3 (WFC3) on the Hubble Space Telescope (HST). The nature of this source is still unclear and could be a stellar object, a light echo, or the very late-time emission powered by some unknown process.

Environmental analysis is a powerful method to reveal the origin of astronomical transients; their progenitors' properties can be inferred from their parental stellar populations in the environment, since most stars are born in groups and stars in each group has very similar ages and chemical abundances \citep[e.g.][]{Anderson2012, K2013a, K2013b, K2018, Maund2016, Maund2017, Maund2018, Sun2020a, Sun2021, Sun2022a}. For AT2018cow, \citet{M2019} used data from the Atacama Large Millimeter/submillimeter Array (ALMA) to show that its host galaxy is rich in molecular gas and has active star formation. \citet{Lyman2020} analysed the integral-field-unit (IFU) data from the Multi-Unit Spectroscopic Explorer (MUSE) mounted on the Very Large Telescope (VLT); they found that AT2018cow's environment is very similar to those of core-collapse SNe in terms of metallicity and H$\alpha$ equivalent width. Both works support AT2018cow arising from a massive stellar explosion; however, more detailed studies are still needed to better constrain the properties of its progenitor.

In this paper, we carry out a comprehensive environmental analysis of AT2018cow. We use a combined dataset from ALMA, VLT/MUSE and HST/WFC3 to probe the molecular gas, ionized gas and stellar populations in its environment and, with a variety of techniques, to derive their accurate parameters. We also try to identify the relationship between AT2018cow and the environmental components. Our aims are to constrain the properties of AT2018cow's progenitor and to explore the nature of its mysterious late-time brightness.

Throughout this paper, we adopt a redshift of 0.01406 and a distance of 63~Mpc for the host galaxy, CGCG~137-068, and a Galactic extinction of $A_V^{\rm mw}$ = 0.24~mag \citep{galebv.ref} and an explosion date of MJD = 58284.79 for AT2018cow (all consistent with \citealt{Xiang2021}). This paper is structured as follows. Section~\ref{data.sec} describes the data source and reduction, and the environmental analysis is carried out in Section~\ref{env.sec}. Section~\ref{implications.sec} discusses the implications for AT2018cow's progenitor and late-time brightness. We finally close this paper with a summary and our conclusions.

\section{Data}
\label{data.sec}

\subsection{ALMA observations}
\label{alma.sec}

AT2018cow was observed by the ALMA 12-m array in the project "The origin of the non-thermal emission of the peculiar transient AT2018cow" (ID: 2017.A.00045.T; PI: Steve Schulze); although the original purpose was to measure the continuum flux of AT2018cow, the CO~(1--0) line (with a rest frequency of 115.271202~GHz) was also covered in one of the 2-GHz spectral windows in the observations on June 30, July 16, August 11 and September 21, 2018. These observations used 43--47 antennas in configurations with maximum baselines of $\sim$300--1400~m, and the on-source integration time was 13--20~min for each execution. The original channel width is 15.625~MHz, corresponding to a velocity resolution of $\sim$40~km~s$^{-1}$. The flux density scale was established using scans of the standard calibrator J1550+0527, and the flux calibration uncertainty was $\sim$3\% for ALMA Cycle-5 band-3 observations \citep{alma.ref}. The phase and water vapour were further checked by observing the nearby calibrator of J1619+247.

We retrieved the data from the ALMA science archive\footnote{\url{https://almascience.nrao.edu/}} and performed calibration with the standard pipeline and the \textsc{casa} package\footnote{\url{https://casa.nrao.edu/}}. The underlying dust continuum emission was subtracted in the uv-plane, and we only considered the frequency range overlapping with our science goals for the final combination of multi-execution data. We made the line datacube from the combined calibrated data using the \textsc{tclean} task with \texttt{Briggs} weighting and \texttt{robust = 0.5} (in order to optimize the sensitivity per frequency bin and the resolution of the final map) and with \texttt{auto-multithresh} mask and \texttt{sidelobethreshold = 2.0}, \texttt{noisethreshold = 4.25}, \texttt{minbeamfrac = 0.3}, \texttt{lownoisethreshold = 1.5}, \texttt{negativethreshold = 15.0} and \texttt{threshold = 3$\sigma$}. The synthesized beam size is 1$\farcs$362 $\times$ 1$\farcs$135, corresponding to 416~pc $\times$ 346~pc at the host galaxy's distance. The noise level in a 15.625-MHz channel is 1.87 $\times$ 10$^{-4}$~mJy~beam$^{-1}$, and the moment-0 map (i.e. the velocity-integrated intensity) has a sensitivity of 0.027~Jy~beam$^{-1}$~km~s$^{-1}$.

Our derived CO~(1--0) map (Fig.~\ref{image.fig}) is slightly different from that of \citet{M2019}. We noted that they have only used the observations performed on June 30 and July 16, 2018 with shorter baselines and poorer spatial resolutions. By including the additional data acquired on August 11 and September 21, 2018, we can achieve a higher spatial resolution and sensitivity.

\subsection{VLT/MUSE observations}
\label{vlt.sec}

AT2018cow was also observed by VLT/MUSE in May 2019 (i.e. $t$ $\sim$1~yr) as part of the All-weather MUse Supernova Integral field Nearby Galaxies (AMUSING) survey IX: supernova and tidal disruption event rates as a function of environment age and metallicity (Program ID: 0103.D-0440; PI: Anderson J.). The total exposure time was 2805s, divided into four exposures with slight offsets and rotations in 90-deg steps to minimise detector artefacts. The field of view is 1$\arcmin$ $\times$ 1$\arcmin$ (wide-field mode) with a sampling of 0$\farcs$2 and the spatial point spread function (PSF) has a full width at half maximum (FWHM) of 0$\farcs$86 (as determined by \citealt{Lyman2020} using isolated stars in the white-light image). The wavelength range is 4750--9350~\AA\ with a 1.25-\AA\ sampling and the spectral resolution ($R$ = d$\lambda$/$\lambda$) varies from $\sim$1770 at the blue end to $\sim$3590 at the red end.

Datacube reduced by the standard pipeline was retrieved from the ESO data archive\footnote{\url{http://archive.eso.org/}} and then corrected for a flux scaling factor of 0.35 from the photometric re-calibration of AMUSING Data Release 1 \citep{Lyman2020}. The \textsc{ifuanal} package \citep{Lyman2018, Lyman2020} was applied for further processing. The datacube was firstly de-redshifted, corrected for Galactic extinction, and masked for bright foreground stars. The spaxels were then binned with Voronoi tessellation to achieve signal-to-noise ratios of at least 30 in the wavelength range of 5590--5680~\AA. The spectrum of each bin was fitted with the \textsc{starlight} package \citep{starlight.ref} based on the \citet{bc03.ref} simple stellar population models, and the fitted stellar continuum was scaled to match and removed from the observed spectrum of the each individual spaxel inside the bin. In this way, we can clean the stellar radiation from the datacube, leaving only nebular emission lines from the ioinzed gas.

\subsection{HST/WFC3 observations}
\label{hst.sec}

\begin{table}
\caption{HST/WFC3 observations.}
\begin{tabular}{cccc}
\hline
\hline
Program & Epoch$^{\rm d}$ & Filter & Exposure \\
ID & (day) & & Time (s) \\
\hline
15600$^{\rm a}$
& 52 & F225W &  770  \\
15600
& 57 & F225W &  770  \\
15600
& 62 & F225W &  770 \\
15974$^{\rm b}$
& 714 & F225W &  1116 \\
& 714 & F336W &  1116 \\
& 714 & F555W &  1044 \\
& 714 & F814W &  1044 \\
16179$^{\rm c}$
& 1136 & F555W &  710 \\
& 1136 & F814W &  780 \\
\hline
\end{tabular} \\
PIs: (a) Foley R.; (b) Levan A.; (c) Filippenko A. \\
(d) Epoch is relative to an explosion date of MJD~= 58,284.79 as estimated by \citet{Xiang2021}.
\label{obs.tab}
\end{table}

Table~\ref{obs.tab} lists the HST data used in this work. They are from three different programs (IDs: 15600, 15974 and 16179), all performed with the Ultraviolet-Visible (UVIS) channel of WFC3. The first program was conducted at $t$ = 52, 57 and 62~d, aiming to observe AT2018cow itself in four UV filters (F218W, F225W, F275W and F336W). In this work, we only use the F225W images from this dataset since the other ones are too shallow to probe any stellar populations efficiently in the environment. The second program was observed at a much later epoch of $t$ = 714~d and covered a large wavelength range with the F225W, F336W, F555W and F814W filters. Lastly, the third program obtained optical F555W and F814W images at $\sim$3~yrs after explosion.

The images were retrieved from the Mikulski Archive for Space Telescopes\footnote{\url{https://archive.stsci.edu/}}. We aligned them with \textsc{tweakreg}, acquiring sub-pixel accuracies, and re-drizzled them with \textsc{astrodrizzle}\footnote{\url{http://drizzlepac.stsci.edu/}}, using \texttt{driz\_cr\_grow = 3} for better cosmic ray removal (all other parameters were left unchanged as in the standard calibration pipeline). We then performed PSF photometry on the images with the \textsc{dolphot} package \citep{dolphot.ref}. In doing this, we used parameters of \texttt{FitSky~= 2}, \texttt{img\_RAper~= 3}, and \texttt{Force1~= 1}, which can better resolve the sources in crowded fields; aperture correction was turned off with \texttt{ApCor~= 0} (typical values are only of several hundredths magnitude, much smaller than the random errors); the other parameters were the same as recommended in the user manual. Artificial star tests were used to determine the detection limits and estimate the additional photometric uncertainties due to crowding.  Throughout this paper all magnitudes are reported in the Vega system.

\subsection{Relative astrometry}
\label{align.sec}

We used 8 common stars to align the MUSE datacube with the HST images, achieving an accuracy of $\sim$0.9~WFC3/UVIS pixel or $\sim$0.2 MUSE spaxel. The ALMA datacube was aligned with the MUSE datacube and HST images using the galaxy's centre as reference; the typical error is 1-2 MUSE spaxels, much smaller than the spatial extents of the molecular or ionized gas.

\section{Environment}
\label{env.sec}

\begin{figure*}
\centering
\includegraphics[width=0.95\linewidth, angle=0]{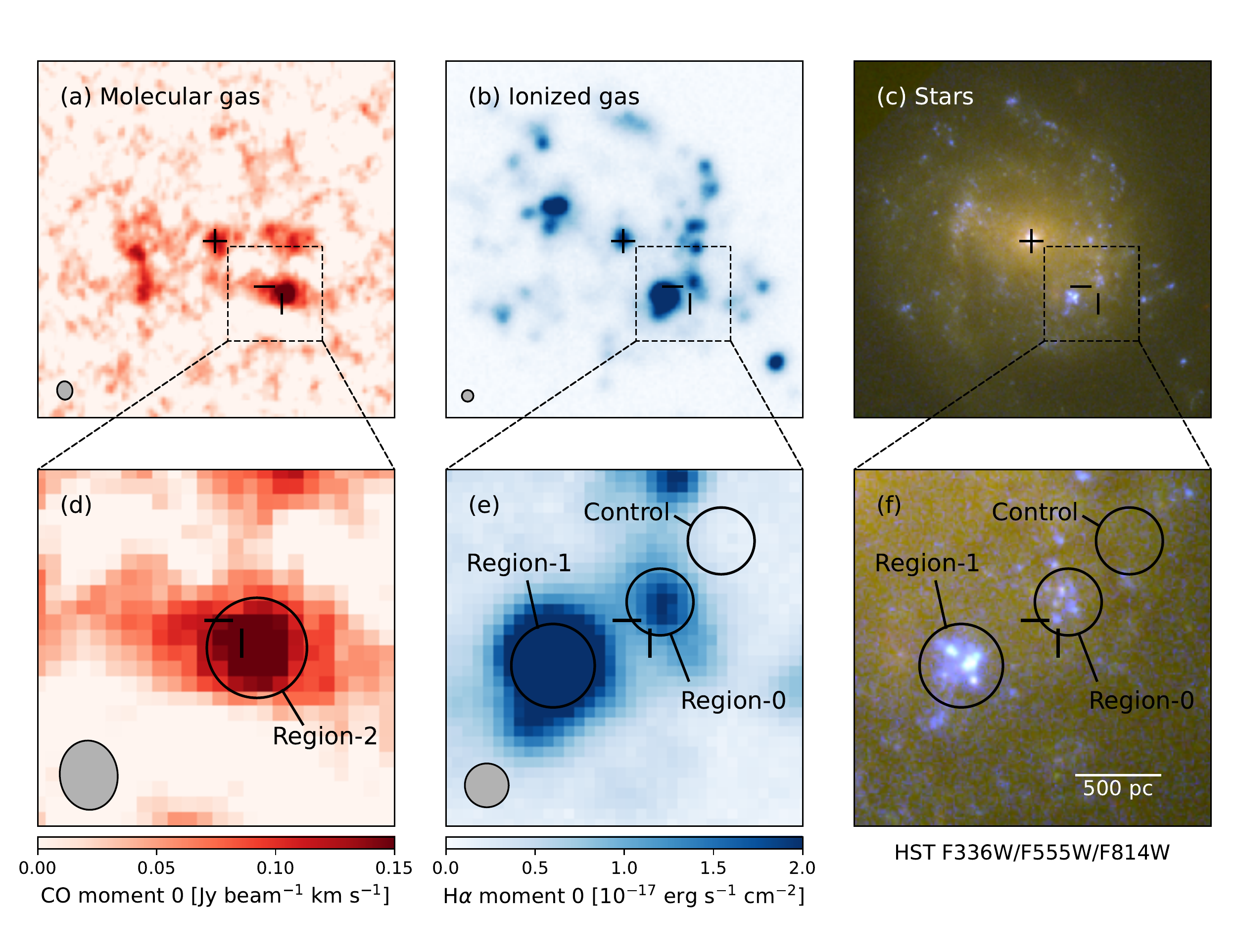}
\caption{AT2018cow's host galaxy (upper panels) and local environment (lower panels): (a, d) velocity-integrated intensity maps of the CO~(1--0) line emission as observed by ALMA; (b, e) wavelength-integrated flux maps of the H$\alpha$ line emission as observed by VLT/MUSE; and (c, f) the F336W/F555W/F814W three-colour composite images as observed by HST/WFC3 at $t$ = 714~d. The crosshairs correspond to the position of AT2018cow, and the "+" symbols in the upper panels indicate the centre of the host galaxy. The spatial resolutions of the CO and H$\alpha$ maps are shown by the grey ellipses/circles in the lower-left corners. The regions defined for environmental analysis are shown as the circles in the lower panels. All images are aligned with north up and east to the left.}
\label{image.fig}
\end{figure*}

\begin{figure*}
\centering
\includegraphics[width=1\linewidth, angle=0]{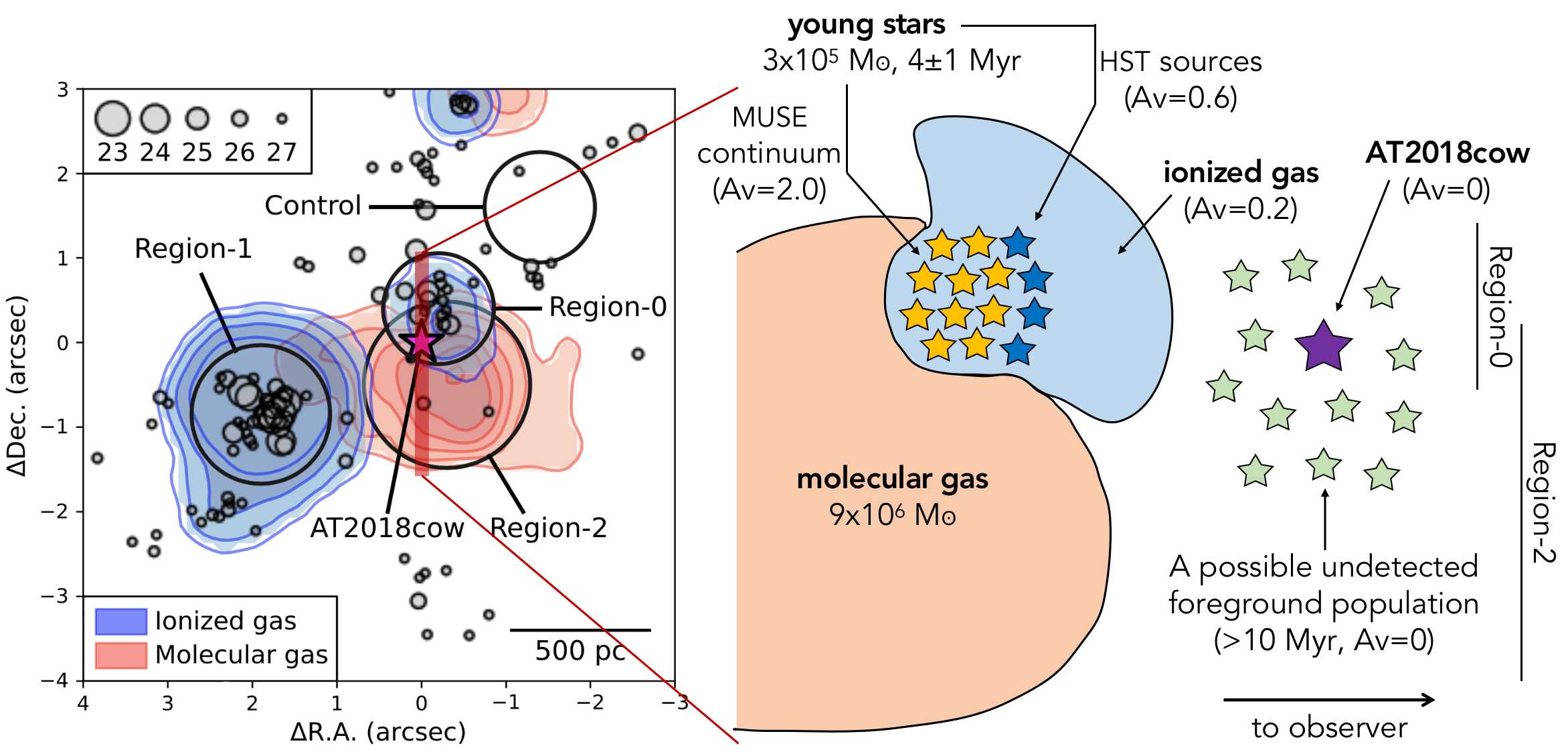}
\caption{Left: the local environment of AT2018cow; blue contours: CO~(1--0) velocity-integrated intensities from 3$\sigma$ to 10$\sigma$ in steps of 1$\sigma$, where $\sigma$ = 0.027 Jy~beam$^{-1}$~km~s$^{-1}$; red contours: H$\alpha$ wavelength-integrated fluxes in levels of 1.2, 1.5, 2.0 and 3.0 $\times$ 10$^{-17}$ erg~s$^{-1}$~cm$^{-2}$; points: sources detected on the HST F555W image with symbol sizes reflecting their magnitudes (according to the legend shown in the upper-left corner); the star in magenta indicates the position of AT2018cow and the large circles in black show the regions defined for environmental analysis. Right: A simple model of the environment of AT2018cow along a vertical slice through its position and the line of sight.}
\label{struct.fig}
\end{figure*}

\begin{table*}
\caption{Parameters derived from the environmental analysis}
\begin{tabular}{lrrrll}
\hline
\hline
Molecular gas & & & \textit{Region-2} & Unit & Note \\
\hline
CO (1-0) velocity-integrated \\
\ \ \ \  line intensity $F_{\rm CO (1-0)}$ & & & 0.15 $\pm$ 0.02 & Jy km s$^{-1}$ \\
CO (1-0) line central velocity $v_{\rm CO (1-0)}$ & & & 4207 $\pm$ 4 & km s$^{-1}$ & Barycentric frame \\
CO (1-0) line width $\sigma_{v, \rm CO (1-0)}$ & & & 33 $\pm$ 4 & km s$^{-1}$\\
CO (1-0) line luminosity $L_{\rm CO (1-0)}$ & & & 1.4 $\pm$ 0.3 & $\times$~10$^{6}$ K km s$^{-1}$ pc$^2$ \\
Molecular gas mass $M_{\rm molecular}$ & & & 6 $\pm$ 1 & $\times$~10$^{6}$ $M_\odot$ & \\
\hline
\hline
Ionized gas & \textit{Region-0} & \textit{Region-1} & & Unit & Note \\
\hline
H$\alpha$ line wavelength-integrated flux $F_{{\rm H}\alpha}$ & 0.52 $\pm$ 0.03 & 2.56 $\pm$ 0.07 & & $\times$~10$^{-15}$ erg s$^{-1}$ cm$^{-2}$\\
H$\alpha$ line central velocity $v_{{\rm H}\alpha}$ & 4194.8 & 4180.1 & & km s$^{-1}$ & Barycentric frame \\
H$\alpha$ line width $\sigma_{v, {\rm H}\alpha}$ & 18.2 & 17.6 & & km s$^{-1}$ & \\
Total extinction $A_V^{\rm tot}$ & 0.46 $\pm$ 0.07 & 0.44 $\pm$ 0.02 & & mag & Corresponding to an internal \\
 & & & & & \ \ \ \ extinction of $A_V^{\rm int}$ $\sim$ 0.2~mag \\
H$\alpha$ line luminosity $L_{{\rm H}\alpha}$ & 0.34 $\pm$ 0.05 & 1.64 $\pm$ 0.05 & & $\times$~10$^{39}$ erg s$^{-1}$ & \\
Electron density $n_e$ & $\lesssim$10 & $\lesssim$10 & & cm$^{-3}$ & \\
Electron temperatrue $T_e$ & $<$10500 & $<$6700 & & K & \\
Oxygen abundance 12 + log(O/H) & $8.44 \pm 0.18$ & $8.43 \pm 0.18$ & & dex & From the strong-line diagnostics \\
& $8.54 \pm 0.11$ & $8.53 \pm 0.11$ & & dex & From photoionisation calculation \\
Nitrogen abundance 12 + log(N/H) & $7.59 \pm 0.09$ & $7.61 \pm 0.10$ & & dex & \\
Sulphur abundance 12 + log(S/H) & $6.95 \pm 0.09$ & $6.96 \pm 0.10$ & & dex & \\
Ionisation parameter log($U$) & $-3.16 \pm 0.05$ & $-2.90 \pm 0.07$ & & dex & \\
Log-age of the ionising \\
 \ \ \ \ sources log($t$/yr) & 6.63$^{+0.09}_{-0.12}$ & $6.26 \pm 0.07$ & & dex & \\
\hline
\hline
Resolved stars & \textit{Region-0} & \textit{Region-1} & & Unit & \\
\hline
Mean log-age log($t$/yr) 
& 6.58 $\pm$ 0.07 & 6.39 $\pm$ 0.01 & & dex & \\
Log-age dispersion $\sigma_{{\rm log(}t/{\rm yr)}}$
& 0.21$_{-0.04}^{+0.07}$ & 0.01 $\pm$ 0.01 & & dex & \\
Mean internal extinction $A_V^{\rm int}$
& 0.6 $\pm$ 0.1 & 0.45 $\pm$ 0.07 & & mag & \\
Internal extinction dispersion $\sigma_{A_V^{\rm int}}$
& 0.05$_{-0.02}^{+0.09}$ & 0.24$_{-0.04}^{+0.07}$ & & mag & \\
\hline
\hline
Possible star clusters & & \textit{Region-1} & & Unit & \\
\hline
Internal extinction $A_V^{\rm int}$ & & $\sim$0.7 & & mag & \\
Log-mass log($M$/$M_\odot$) & & $\sim$4.7 & & dex & \\
\hline
\hline
Stellar continuum & \textit{Region-0} & \textit{Region-1} & & Unit & \\
\hline
Internal extinction $A_V^{\rm int}$ & $\sim$2.0 & $\sim$1.5 & & mag & \\
Total stellar mass $M_{\rm stellar}$ & $\sim$3 & $\sim$3 & & $\times$ 10$^5$~$M_\odot$ & \\
\hline
\end{tabular} \\
\label{params.tab}
\end{table*}

Figure~\ref{image.fig} shows the maps of molecular gas, ionized gas and stellar populations in the host galaxy of AT2018cow, traced by CO~(1--0) observed by ALMA, H$\alpha$ observed by VLT/MUSE and UV-optical images acquired by HST/WFC3. The galaxy has a bright nuclei and very flocculent spiral arms. The arms are only marginally depicted by the young stars and H~\textsc{ii} regions, and can hardly be seen in the spatial distribution of molecular gas. Star formation inside such a flocculent galaxy is most likely self-propagating instead of being controlled by spiral density waves.

Figure~\ref{struct.fig} shows the structure of AT2018cow's local environment. In its vicinity, there is a very prominent concentration of molecular gas associated with two giant star-forming complexes. Both complexes are very populous in young massive stars/star clusters, which have photoionized the surrounding gas and given rise to the bright H~\textsc{ii} regions. AT2018cow is spatially coincident with one complex; at the time of the HST observations, AT2018cow was still bright in all four filters (F225W, F336W, F555W and F814W) although the nature of this late-time source remains mysterious \citep{Sun2022b}.

We define four circular regions for the next environmental analysis. \textit{Region-0} is centred on the star-forming complex associated with AT2018cow (following the convention of \citealt{Lyman2020}) and has a radius of 200~pc. \textit{Region-1} has a larger radius of 250~pc and encloses the other star-forming complex. With a radius of 300~pc, \textit{Region-2} is used to analyse the molecular gas. We further define a control region with the same radius as \textit{Region-0}, located at a similar galactocentric distance but without obvious star formation, in order to estimate the light from older stellar populations (which will be used in Section~\ref{continuum.sec}).

\subsection{Molecular gas}
\label{molecular.sec}

\begin{figure}
\centering
\includegraphics[width=0.9\linewidth, angle=0]{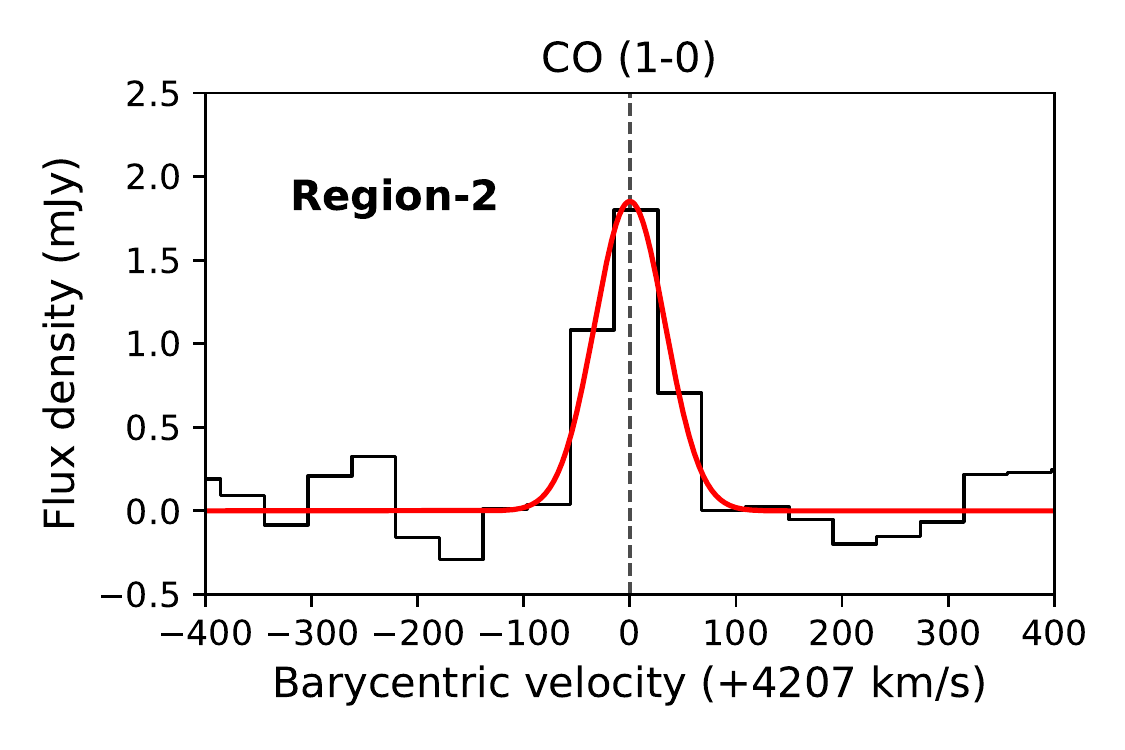}
\caption{Black: continuum-subtracted ALMA CO~(1--0) spectrum integrated over \textit{Region-2}; the velocity is in the Barycentric frame and with respect to the line's rest frequency of 115.271202~GHz plus 4207~km~s$^{-1}$. Red: Gaussian-profile fit to the data; the vertical dashed line shows the central velocity derived from the fitting.}
\label{CO.fig}
\end{figure}

Figure~\ref{CO.fig} shows the ALMA spectrum integrated over \textit{Region-2}; the continuum has been subtracted by fitting with a zeroth-order polynomial. The CO~(1--0) line is clearly detected, and with a Gaussian fit we derive a velocity-integrated intensity of $F_{\rm CO (1-0)}$ = 0.15 $\pm$ 0.02~Jy~km~s$^{-1}$, a central velocity of 4207 $\pm$ 4~km~s$^{-1}$ (in the barycentric frame) and a line width (standard deviation) of 33 $\pm$ 4~km~s$^{-1}$. The measured parameters are then converted to a luminosity of $L_{\rm CO (1-0)}$ = 1.4 ($\pm$~0.3) $\times$~10$^{6}$~K~km~s$^{-1}$~pc$^2$ with
\begin{multline}
\left(\dfrac{L_{\rm CO (1-0)}}{\rm K\ km\ s^{-1}\ pc^2}\right) = \\
\left( \dfrac{3.25 \times 10^7}{(1+z)^3} \right)
\left(\dfrac{F_{\rm CO (1-0)}}{\rm Jy\ km\ s^{-1}}\right)
\left(\dfrac{\nu_{\rm obs}}{\rm GHz} \right)^{-2}
\left(\dfrac{D}{\rm Mpc} \right)^2,
\end{multline}
where $z$ is the redshift, $v_{\rm obs}$ the observed line frequency, and $D$ the distance to the target. By assuming a CO-to-H$_2$ conversion factor\footnote{Note that the CO-to-H$_2$ conversion factor does depend on metallicity and the value of \citet{Utomo2015} is for the molecular clouds in the Milky Way. Although the environment of AT2018cow has a slightly sub-solar abundance of 12 + log(O/H) = 8.44 (see Section~\ref{ionized.sec}), the CO-to-H$_2$ conversion factor at this metallicity is not significantly different \citep{Leroy2011}. The conversion factor becomes significantly larger at 12 + log(O/H) < 8.2 in the metal-poor galaxies such as NGC~6822 and the Small Magellanic Cloud.} of $X_{\rm CO}$ = 2 $\times$~10$^{20}$~cm$^{-2}$~(K~km~s$^{-1}$)$^{-1}$ \citep{Utomo2015}, we estimate a molecular gas mass of 6 ($\pm$~1) $\times$~10$^6$~$M_\odot$ (including hydrogen and helium), using the relation
\begin{equation}
\left(\dfrac{M_{\rm molecular}}{M_\odot}\right) = 4.4 \times \left(\dfrac{L_{\rm CO (1-0)}}{\rm K\ km\ s^{-1}\ pc^2}\right).
\end{equation}

These results support \textit{Region-2} being a very massive agglomeration of molecular gas. Its mass is almost twice that of LMC's prominent "Molecular Ridge" (with 30~Dor on its northern tip; \citealt{Ott2008}) and even comparable to some of the mini-starburst molecular cloud complexes \citep[e.g.][]{N16}. This large reservoir of cold molecular gas provides enough material for the active star formation seen in the environment of AT2018cow.

\subsection{Ionized gas}
\label{ionized.sec}

\begin{figure*}
\centering
\includegraphics[width=0.85\linewidth, angle=0]{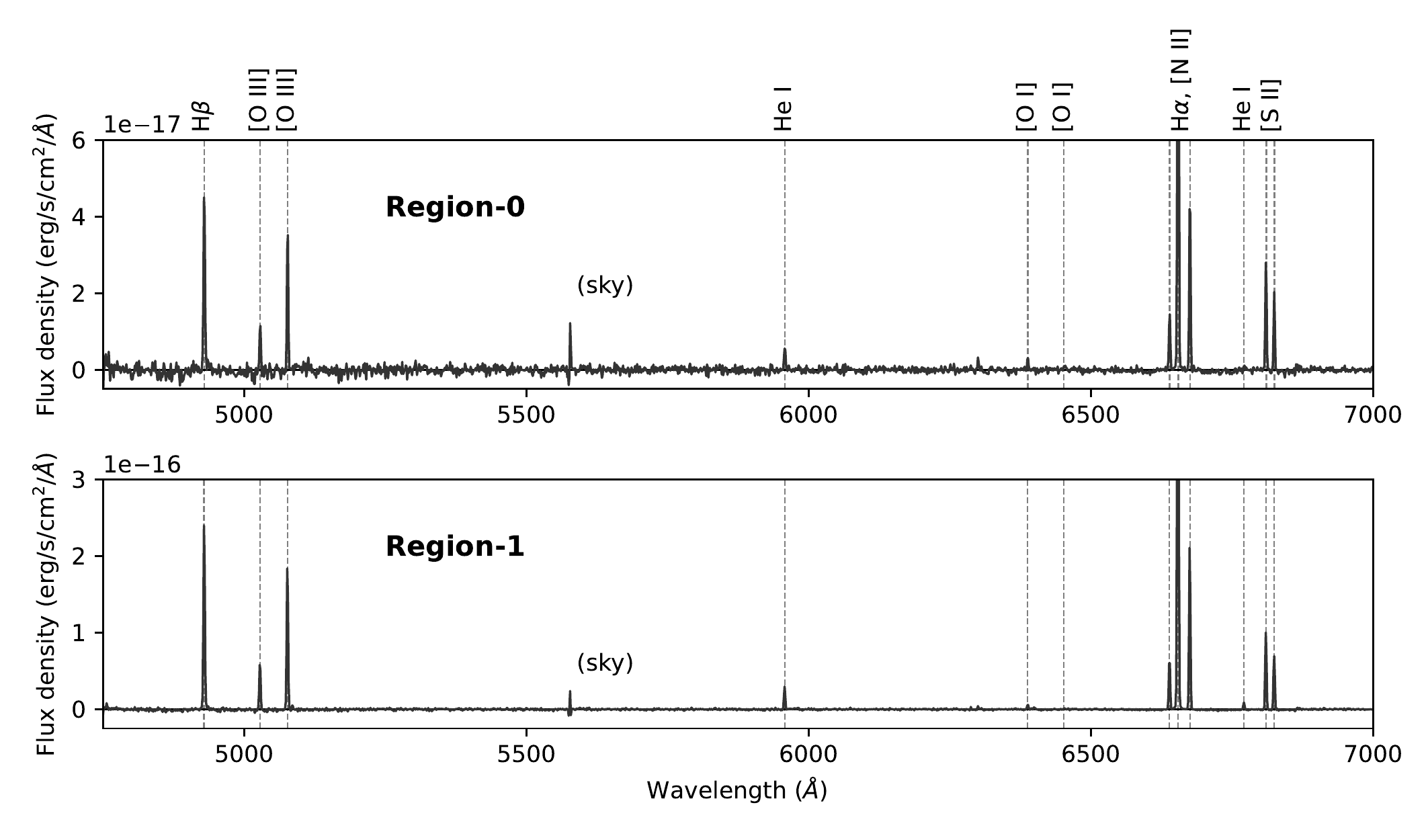}
\caption{Continuum-subtracted MUSE spectra integrated over \textit{Region-0} (upper) and \textit{Region-1} (lower); the red part at $>$7000~\AA\ is not displayed. The vertical dashed lines label the most prominent nebular lines emitted from the ionized gas.}
\label{spec.fig}
\end{figure*}

\begin{figure*}
\centering
\includegraphics[width=1\linewidth, angle=0]{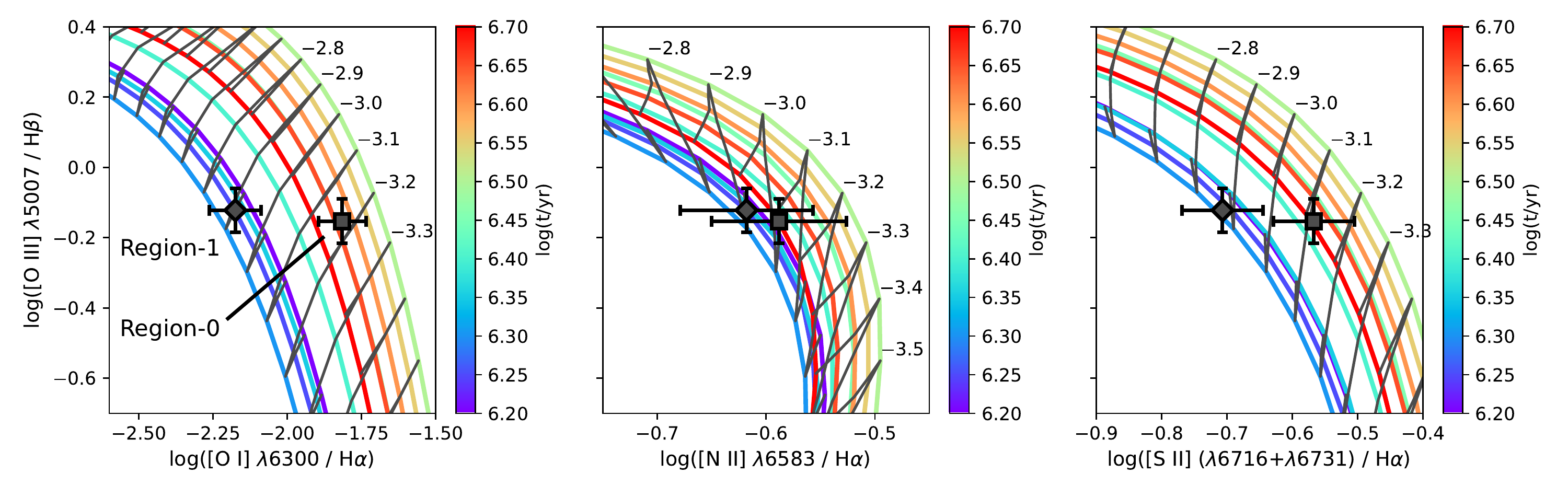}
\caption{BPT(-like) diagrams of the nebular emission-line ratios. The square and diamond symbols show the values measured for \textit{Region-0} and \textit{Region-1}, respectively, and the error bars reflect their measurement uncertainties. The solid lines are theoretical values predicted by photoionization calculation (with chemical abundances matching the best-fitting values). The grey lines are loci with varying ionising source ages and fixed ionisation parameters (values of log$U$ are labeled next to the lines), while the coloured lines have fixed ionising source ages (according to the colour bar) and varying ionisation parameters. Note that the relation is not always monolithic between the emission-line ratios and the ionising source ages.}
\label{bpt.fig}
\end{figure*}

Figure~\ref{spec.fig} shows the MUSE spectra integrated over \textit{Region-0} and \textit{Region-1} from the continuum-subtracted datacube. A series of nebular emission lines are clearly revealed, including the recombination lines of H$\alpha$, H$\beta$, He~\textsc{i}~$\lambda$5876~\AA\ and He~\textsc{i}~$\lambda$6678~\AA, and the collisionally-excited lines from nitrogen ([N~\textsc{ii}]~$\lambda\lambda$6548, 6583), oxygen ([O~\textsc{i}]~$\lambda\lambda$6300, 6363, [O~\textsc{iii}]~$\lambda\lambda$4959, 5007), and sulphur ([S~\textsc{ii}]~$\lambda\lambda$6716, 6731). We fit Gaussian profiles to the lines to determine their wavelength-integrated fluxes, central velocities and widths; we also remove the instrumental broadening \citep[using the line spread function of][]{Guerou2017} from the measured line widths to acquire the intrinsic velocity dispersions. Values obtained for H$\alpha$ are listed in Table~\ref{params.tab}.

\subsubsection{Extinction}

With Balmer decrement we determine total extinctions of $A_V^{\rm tot}$ = 0.46 $\pm$~0.07~mag for \textit{Region-0} and 0.44 $\pm$ 0.02 for \textit{Region-1}, assuming an intrinsic flux ratio of H$\alpha$/H$\beta$ = 2.87 \citep{agn2.ref} and a standard extinction law with $R_V$ = 3.1 \citep{F04.ref}. By subtracting the Galactic extinction (0.24~mag), these values correspond to an internal extinction of $A_V^{\rm int}$ $\sim$ 0.2~mag for the ionized gas in both regions, arising from dust within the host galaxy.

\subsubsection{H$\alpha$ luminosity}

Using the above derived parameters we calculate the extinction-corrected luminosities of H$\alpha$ for the two regions. Large values of $L$(H$\alpha$) = 3.4~($\pm$~0.5) $\times$~10$^{38}$~erg~s$^{-1}$ and 1.64~($\pm$~0.05) $\times$~10$^{39}$~erg~s$^{-1}$ are obtained for \textit{Region-0} and \textit{Region-1}, respectively, with the latter even comparable to that of the 30~Dor mini-starburst region (3.9 $\times$ 10$^{39}$~erg~s$^{-1}$; \citealt{K86.ref}) within a factor of 2.

\subsubsection{Electron density and temperature}

The [S~\textsc{ii}]~$\lambda$6716/$\lambda$6731 line ratios (1.49 for \textit{Region-0} and 1.44 for \textit{Region-1}) suggest that both regions have very low electron densities of $n_e$ $\lesssim$ 10~cm$^{-3}$ \citep{agn2.ref}. The emission-line ratios of [O~\textsc{i}]~($\lambda$6300+$\lambda$6363)/$\lambda$5577 and [N~\textsc{ii}]~($\lambda$6548+$\lambda$6583)/$\lambda$5755 can be used to estimate electron temperature in ionized nebulae. Unfortunately, neither [O~\textsc{i}]~$\lambda$5577 nor [N~\textsc{ii}]~$\lambda$5755 are detected in the spectra; the detection limits constrain the electron temperature to be $T_e$ $<$~10500~K and $<$~6700~K for \textit{Region-0} and \textit{Region-1}, respectively.

\subsubsection{Oxygen abundance}

We use the strong-line method to obtain a rough estimate of the oxygen abundance with the O3N2 calibration of \citet{o3n2.ref}. 12 + log(O/H) = 8.44 (\textit{Region-0}) and 8.43 (\textit{Region-1}) are obtained, which are very close to each other. These values are slightly lower than those derived by \citet{Lyman2020} (8.57 and 8.66 based on a different calibration of \citealt{Dopita2016}), and we note that the strong-line method is subject to relatively large uncertainties (typically $\sim$0.18~dex). Our results suggest slightly sub-solar abundances of [O/H] $\sim$ $-$0.25 adopting an absolute solar scale of 12 + log(O/H)$_\odot$ = 8.69 \citep{solar.ref}.

\subsubsection{Ionising sources}
\label{calculation.sec}

We further carry out a detailed calculation of photoionisation for \textit{Region-0} and \textit{Region-1} with the \textsc{cloudy} package \citep[v17.02;][]{cloudy.ref}. The method is very similar to that described in \citet[][see also \citealt{Xiao2019} and \citealt{Sun2020b}]{Sun2021}. In brief, we assume that the ionising radiation comes from a stellar population of a single age log($t$/yr), simulated with the \textsc{bpass}~(v2.1) synthetic spectra including binaries \citep{bpass.ref}, and that the ionized gas is a radiation-bounded shell (i.e. optically thick to the ionising photons). The intensity of the ionizing radiation is characterized by the ionisation parameter
\begin{equation}
U = \dfrac{Q({\rm H})}{4\pi r_0^2 n_{\rm H} c},
\end{equation}
where $Q({\rm H})$ is the luminosity of the hydrogen-ionizing photons, $r_0$ the inner radius of the shell, $n_{\rm H}$ the hydrogen-atom number density (including all atoms in all forms of particles), and $c$ is the speed of light. We assume relative-to-solar abundances of [M/H] = $-$0.25 for all heavy elements (i.e. same as the oxygen abundance estimated by the strong-line method), but allow the nitrogen, oxygen and sulphur abundances to vary around this value with Gaussian-prior probability distributions (with a standard deviation of 0.18~dex, i.e. the typical uncertainty of the strong-line method). We use the default set of absolute solar composition in \textsc{cloudy}, in which the solar nitrogen, oxygen and sulphur abundances are 12 + log(M/H) = 7.93, 8.69 and 7.27 (M = N, O and S), respectively. A hydrogen-atom density of $n_{\rm H}$ = 10~cm$^{-3}$ is used and fixed in the calculation.

We then solve for log($t$/yr), log$U$, [N/H], [O/H] and [S/H] by comparing the observed and predicted emission-line ratios of [O~\textsc{iii}]~$\lambda$5007/H$\beta$,  [O~\textsc{i}]~$\lambda$6300/H$\alpha$, [N~\textsc{ii}]~$\lambda$6583/H$\alpha$ and [S~\textsc{ii}]~($\lambda$6716+$\lambda$6731)/H$\alpha$ (see Fig.~\ref{bpt.fig} for their BPT and BPT-like diagrams). In doing this, we conservatively assume a 10\% uncertainty in the theoretical line fluxes. The results show that \textit{Region-0} has a very young ionising source age of log($t$/yr) = 6.63$^{+0.09}_{-0.12}$ with an ionisation parameter of log$U$ = $-$3.16 $\pm$ 0.05. For \textit{Region-1}, the ionising sources have an even younger age of log($t$/yr) = 6.26 $\pm$ 0.07 and higher ionisation parameter of log$U$ = $-$2.90 $\pm$ 0.07. The nitrogen, oxygen and sulphur abundances are very similar between the two regions (values are all listed in Table~\ref{params.tab}).

\subsection{Stellar populations}
\label{stars.sec}

\begin{figure*}
\centering
\includegraphics[width=0.95\linewidth, angle=0]{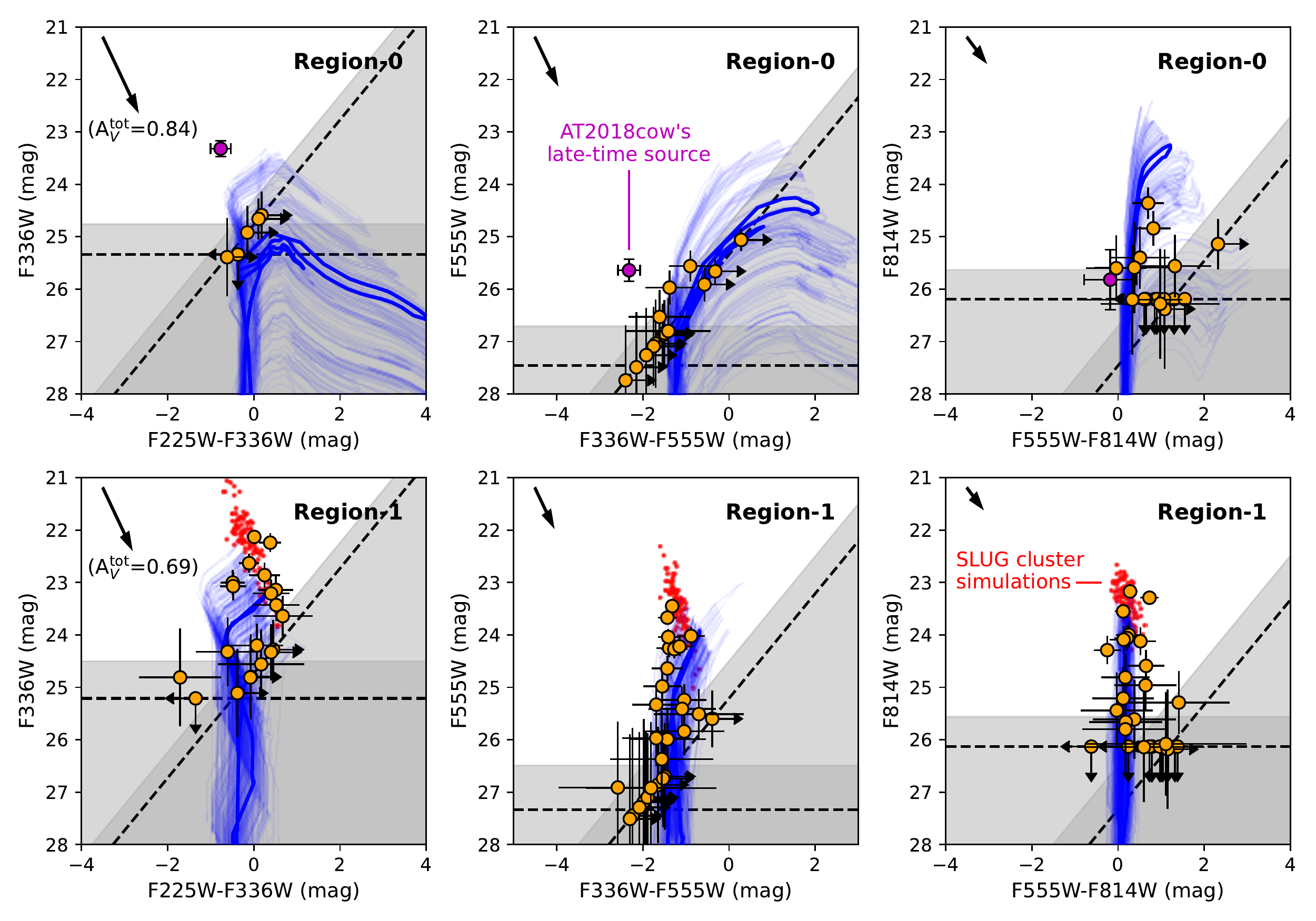}
\caption{CMDs of sources detected on the HST images (orange data points) in \textit{Region-0} (upper panels) and \textit{Region-1} (lower panels). The error bars reflect their 1$\sigma$ photometric uncertainties, the dashed lines show the 50\% detection limits, and the grey-shaded regions show where $\leq$68\% of the artificial stars can be successfully recovered. Model stellar populations are fitted to the observed sources (see text); the thick blue lines are \textsc{parsec} \citep[v1.2S;][]{parsec.ref} single-stellar isochrones corresponding to the mean log-age and mean extinction derived from the fitting, and the thin blue lines are isochrones from 100 random realisations according to the stellar log-age and extinction distributions of the model populations. The red dots in the lower panels are 100 \textsc{slug} simulations of unresolved star clusters with the same mean log-age for \textit{Region-1}. The arrows in the upper-left corners are reddening vectors for a standard extinction law with $R_V$ = 3.1, and their lengths correspond to the total (Galactic + internal) extinctions derived from the model population fitting. For comparison, the late-time source of AT2018cow at $t$ = 714~d \citep[detected by][]{Sun2022b} is also displayed as the magenta data point in the upper panels, with error bars showing its 3$\sigma$ photometric uncertainties.}
\label{cmd.fig}
\end{figure*}

\begin{figure}
\centering
\includegraphics[width=1\linewidth, angle=0]{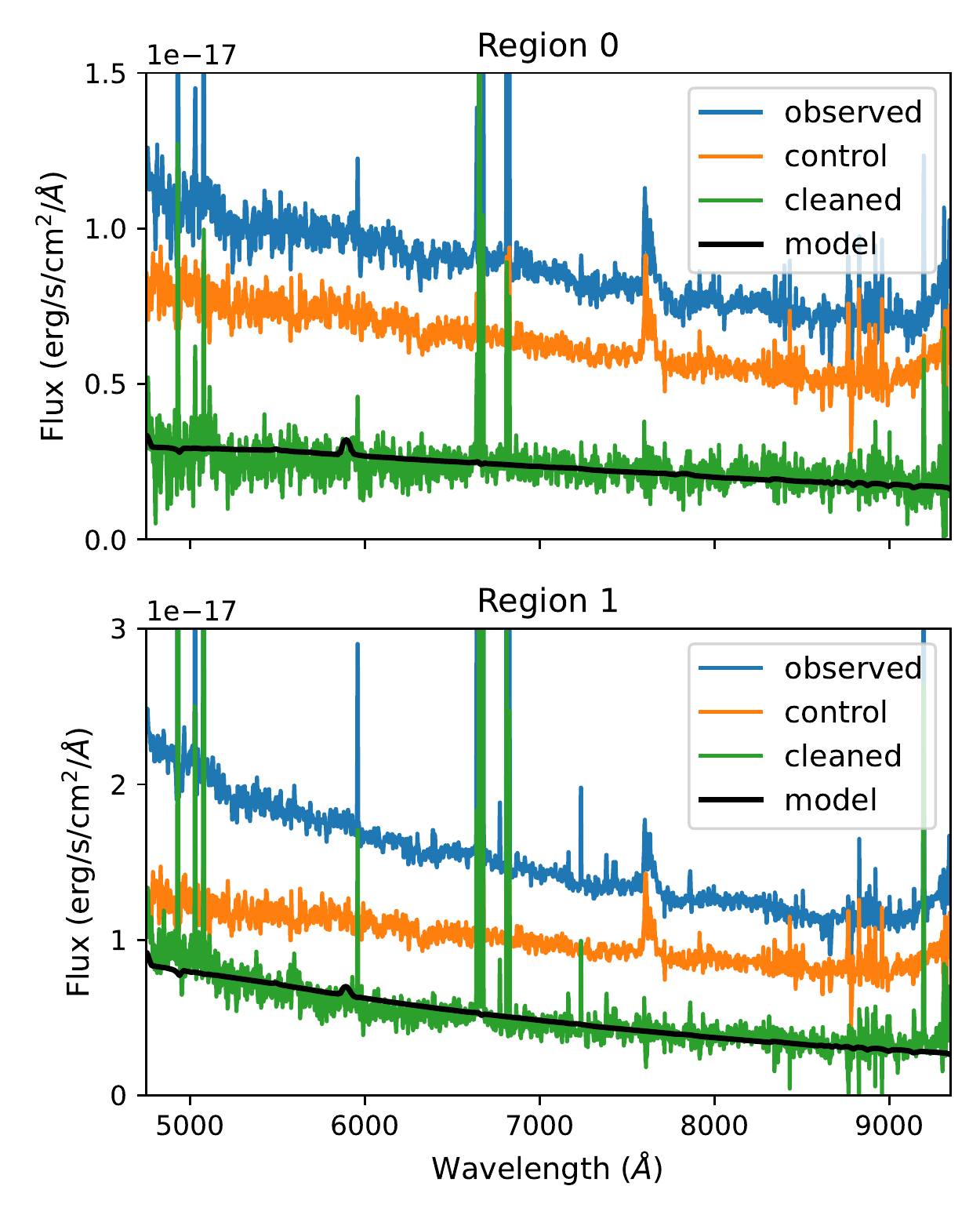}
\caption{The observed (blue), control (orange) and cleaned (green) spectra for \textit{Region-0} (upper panel) and \textit{Region-1} (lower panel). A 5-pixel median filter is applied to smooth the spectra. The black lines are the best-fitting model spectra to the cleaned stellar continuum. In the model spectra, the emission feature at 5808~\AA\ arises from the Wolf-Rayet stars in the population.}
\label{cont.fig}
\end{figure}

\subsubsection{Resolved stars}
\label{resolved.sec}

Figure~\ref{cmd.fig} shows the colour-magnitude diagrams (CMDs) of sources detected on the HST images. With a hierarchical Bayesian approach, we fit model stellar populations to the observed sources in order to derive their accurate parameters. Single stars in the model populations are simulated with the \textsc{parsec} \citep[v1.2S][]{parsec.ref} stellar isochrones, and the binaries are assumed to have composite fluxes equal to the simple sum of their primary and secondary stars\footnote{Binary interaction could be common and important for massive stars, and the product from merger or mass transfer will be more massive and luminous, thus appear to be younger than its true age \citep[e.g.][]{Schneider2014, Stevance2020}. This effect is not included in our modelling, which could possibly lead to an underestimation of the population ages. The ages we derived (see below), however, are very consistent with those from photoionization calculation, in which the interacting binaries have been included (Section~\ref{calculation.sec}). It is possible that the interacting binaries with very high masses, which would appear in the upper part of the CMDs and have the most impact on the fitting, are very rare and might be absent in the analyzed regions due to stochastic sampling.}. The algorithm is detailed in \citet{Maund2016} and \citet{Sun2021} and has been successfully applied in a number of studies \citep[e.g.][]{Maund2017, Maund2018, Sun2022a, Sun2022c}. In this paper, we adopt the same model settings as in those previous works; the only exceptions are (1) only one model population is used to fit each region given the small number of detected sources; and (2) stars from the model population are assumed to have Gaussian distributions in log-age and internal extinction, whose mean values and standard deviations are all left as free parameters. For each region, there are four free parameters to be fitted: the mean log-age log($t$/yr), log-age dispersion $\sigma_{{\rm log(}t{\rm /yr)}}$, mean internal extinction $A_V^{\rm int}$ and the internal extinction dispersion $\sigma_{A_V^{\rm int}}$. We use flat priors for the mean log-age and mean internal extinction; the dispersions are assumed to have uniform prior distributions on logarithmic scales over the ranges of 0 $\leq$ log[$\sigma_{{\rm log(}t{\rm /yr)}}$/(0.01~dex)] $\leq$ 2 and 0 $\leq$ log[$\sigma_{A_V^{\rm int}}$/(0.01~mag)] $\leq$ 2, penalizing large values. The posterior probability distributions are then solved with the \textsc{dynesty} package \citep{dynesty.ref}.

The fitting results are listed in Table~\ref{params.tab} and shown as the isochrones in the CMDs. \textit{Region-0} has a mean log-age of log($t$/yr) = 6.58 $\pm$ 0.07, very close to that from photoionization calculation (6.63$^{+0.09}_{-0.12}$; Section~\ref{calculation.sec}). The mean log-age derived for \textit{Region-1}, log($t$/yr) = 6.39 $\pm$ 0.01, is slightly larger than that from photoionization calculation (6.26 $\pm$ 0.07; Section~\ref{calculation.sec}) but still within 2$\sigma$ uncertainties. These results are obtained by assuming a standard extinction law with $R_V$ = 3.1; changing $R_V$ has very little influence on the derived ages.

On the other hand, the derived mean internal extinctions of $A_V^{\rm int}$ = 0.6 $\pm$ 0.1~mag for \textit{Region-0} and 0.45 $\pm$ 0.07~mag for \textit{Region-1} are larger than that for the ionized gas from Balmer decrement (which is $\sim$0.2~mag after removing a Galactic extinction of 0.24~mag). This difference possibly suggests that the observed stars and the ionized gas could be distributed at different places with respect to the dust along the line of sight (which shall be discussed later).

\subsubsection{Possible star clusters}

Some of the sources in \textit{Region-1} appear (marginally) brighter than the best-fitting isochrones, which could be binary/multiple stars and/or spatially-unresolved star clusters\footnote{At the distance of AT2018cow, the PSF width of HST/WFC3 corresponds to $\sim$20~pc, much larger than the typical sizes of compact star clusters; therefore, it can be very difficult to distinguish stars and star clusters based on source morphologies.}. For comparison, we run 100 simulations of star clusters with the \textsc{slug} package \citep{slug.ref1, slug.ref2}, accounting for the stochastic sampling effect of stellar masses from the initial mass function (IMF). The mean log-age derived by isochrone fitting (Section~\ref{resolved.sec}) is used as the cluster age (fixed), and we assume cluster masses of log($M$/$M_\odot$) = 4.7 $\pm$ 0.1 and internal extinctions of 0.7 $\pm$ 0.3~mag. The simulated clusters are displayed as the red dots in the CMDs (Fig.~\ref{cmd.fig}), which agree well with the very bright sources observed in this region.

\subsubsection{Stellar continuum}
\label{continuum.sec}

Figure~\ref{cont.fig} shows the MUSE spectra integrated over \textit{Region-0} and \textit{Region-1} from the original datacube \textit{without} continuum subtraction. In order to isolate the light from the young star-forming complexes, we use the spectrum from the nearby control field (at a similar galactic radius and without obvious star formation) to estimate the contributions from the older stellar populations. The control spectrum is scaled according to the region areas and subtracted from those of \textit{Region-0} and \textit{Region-1} to obtain cleaned spectra. Ideally, the continuum of the cleaned spectra should arise only from the young stellar populations in these two regions. We then fit the cleaned stellar continuum with synthetic spectra from the \textsc{bpass} (v2.1) binary population models. For each region, we use a single and fixed age\footnote{Note that the spectra of young populations are very close to the Rayleigh-Jeans law over the observed wavelengths, and hence are very insensitive to the assumed age.} as derived in Section~\ref{resolved.sec}, redden the synthetic spectrum with the Galactic and internal extinctions, and scale it to match the data.

We find internal extinctions of $A_V^{\rm int}$ $\sim$ 2.0~mag for \textit{Region-0} and 1.5~mag for \textit{Region-1} are required to obtain reasonable fits of their stellar continuum. These values are much larger than those derived for the HST sources and shall be discussed later. The continuum fits derive a total (initial or current) stellar mass of $\sim$3 $\times$~10$^5$~$M_\odot$ for each of the two regions. This large value is comparable to that of 30~Dor's ionizing cluster \citep[4.5 $\times$ 10$^{5}$~$M_\odot$;][]{Bosch2009} and again confirms that \textit{Region-0} and \textit{Region-1} are very active star-forming regions.



\subsection{A simple model of the environment}
\label{model.sec}

The above analysis shows AT2018cow's local environment being very active in recent star formation. In brief summary, the environment includes: (1) a large concentration of molecular gas with a mass of 6 ($\pm$ 1) $\times$ 10$^6$~$M_\odot$; (2) two active star-forming complexes, each with a stellar mass of $\sim$2 $\times$ 10$^5$~$M_\odot$; and (3) two giant H~\textsc{ii} regions photoionized by the young stars. Based on the photoionization calculation (Section~\ref{ionized.sec}) and fitting of the resolved stars (Section~\ref{stars.sec}), we adopt a final value of log($t$/yr) = 6.6 $\pm$ 0.1 or 4 $\pm$ 1~Myr as the stellar age estimate for \textit{Region-0} (with a conservative error bar accounting for the fitting errors and other possible uncertainties) and $\lesssim$2.5~Myr for \textit{Region-1}. The two regions have very high H$\alpha$ luminosities of 3.4 ($\pm$ 0.5) $\times$~10$^{38}$~erg~s$^{-1}$ and 1.64 ($\pm$ 0.05) $\times$~10$^{39}$~erg~s$^{-1}$, respectively. All these parameters support them to be 30~Dor-like mini-starburst regions.

It is worth noting that the internal extinctions derived for the different components are not the same. For \textit{Region-0}, the value from the MUSE stellar continuum ($A_V^{\rm int}$ = 2.0~mag) is much higher than that from the HST resolved sources ($A_V^{\rm int}$ = 0.6 $\pm$ 0.1~mag). This could be due to a detection bias; while most of the stars may have large extinctions and give rise to a highly reddened continuum, only the most luminous and least extinct stars can be resolved as point sources on the HST images. The different (mean) extinctions may suggest they are located at different places along the line of sight, with most stars with high extinctions hiding behind the dust and the few low-extinction ones statistically closer to the observer\footnote{Note that this statement is only statistically meaningful. The interstellar medium is usually very patchy, and the stars and dust are often intermingled with each other. For an individual star, a lower/higher extinction does not mean it must reside in the foreground/background. For a population of stars, however, those in the foreground/background will statistically have a lower/larger \textit{mean} extinction.}.

The molecular gas should most likely reside in the background of the observed stars and ionized gas, since otherwise its large reservoir of dust would almost completely obscure any UV-optical light. From Balmer decrement we derived an internal extinction of $\sim$0.2~mag for the ionised gas (Section~\ref{ionized.sec}), lower than those for the stellar continuum and resolved sources. This places the nebular line-emitting region in the foreground of the stars so that the lines are absorbed by a smaller amount of dust. It is likely that the H~\textsc{ii} region ionized by the newly formed stars has broken out from the molecular gas and is expanding outward to the observer (i.e. the champagne flow model; \citealt{flow.ref}). Figure~\ref{struct.fig} (right) shows a schematic plot of the various components in this simple model of AT2018cow's local environment.

\section{Implications}
\label{implications.sec}

\subsection{The progenitor of AT2018cow}

\begin{figure*}
\centering
\includegraphics[width=0.95\linewidth, angle=0]{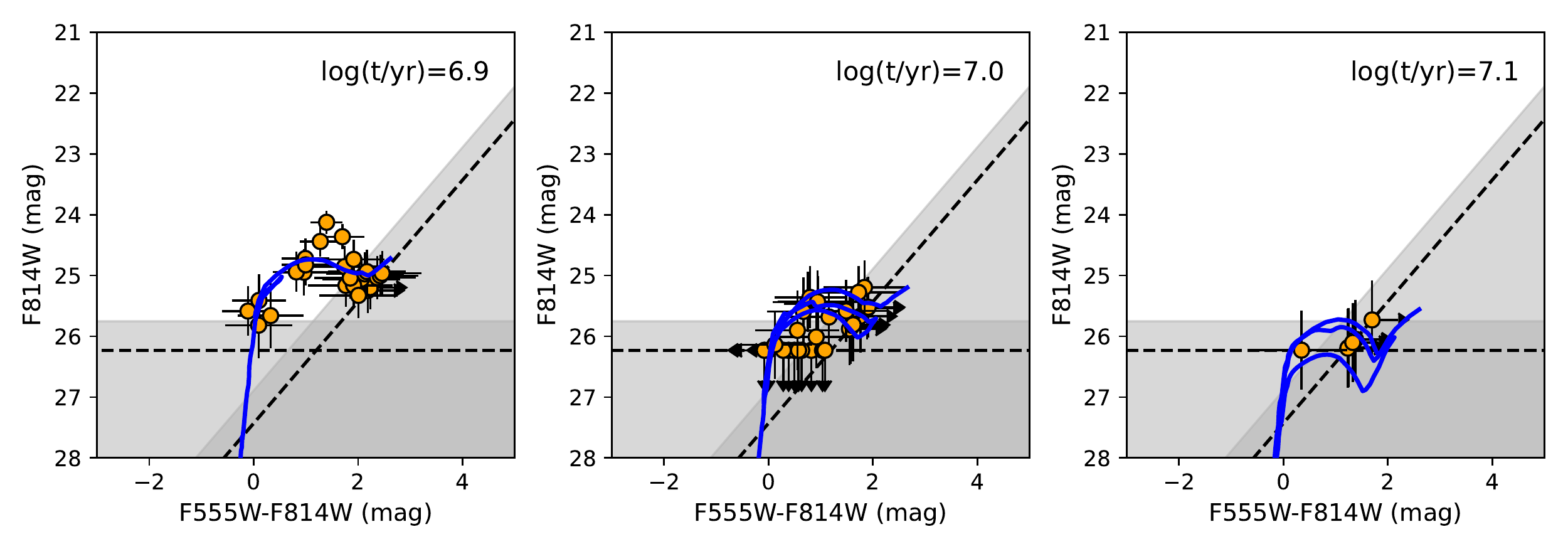}
\caption{CMDs of simulated stellar populations (orange data points) with different ages of log($t$/yr) = 6.9, 7.0 and 7.1 (from left to right). The error bars reflect their 1$\sigma$ photometric uncertainties, the dashed lines show the 50\% detection limits, and the grey-shaded regions show where $\leq$68\% of the artificial stars can be successfully recovered. The thick blue lines are \textsc{parsec} \citep[v1.2S;][]{parsec.ref} single-stellar isochrones. The arrows in the upper-left corners are reddening vectors for a standard extinction law with $R_V$ = 3.1, and their lengths correspond to the Galactic extinction (with no internal extinction from the host galaxy). All three model populations have no detected sources in the F225W and F336W filters, so only the optical CMDs are displayed.}
\label{for.fig}
\end{figure*}

The above environmental analysis allows us to make useful implications on the progenitor of AT2018cow. Its local environment is a typical star-forming region and supports it being the explosion of a massive star (i.e. a young progenitor). It seems unlikely to be a tidal disruption event of a low-mass star (i.e. an old progenitor), since in this case AT2018cow could appear almost anywhere randomly in the galaxy. More precise constraints on the progenitor, however, require a correct identification of its relationship with the environmental components -- in particular, whether the progenitor was a member of, and therefore has the same age as, the mini-starburst in \textit{Region-0}, which AT2018cow is spatially coincident with on the sky plane.

\subsubsection{Scenario 1: the progenitor was a member of the mini-starburst}

If AT2018cow originated from the mini-starburst in \textit{Region-0}, the very young age of 4 $\pm$ 1~Myr would imply a very massive progenitor for this transient. For single-star evolution, this age corresponds to a large initial mass of $M_{\rm ini}$ = 66$^{+47}_{-22}$~$M_\odot$, according to the \textsc{parsec} \citep[v2.1S;][]{parsec.ref} stellar isochrones. The initial mass could also be different if the progenitor has undergone significant binary interaction, such as merger or mass transfer with a companion.

It is worth noting, however, that the internal extinction for AT2018cow itself is close to zero. Its apparent colours at early times ($t$ $<$ 60~d) were very blue ($B - V$ $\sim$ $-$0.1~mag), suggesting negligible extinction from the host galaxy \citep{Kuin2019, Perley2019, Xiang2021}. This is also supported by the very hot SED of AT2018cow's late-time brightness at $t$ = 714~d \citep[reported by][]{Sun2022b} and we find that even a small amount of internal extinction would make a very hot blackbody spectrum redder than observations (especially at the UV wavelengths). Thus, AT2018cow's internal extinction is lower than those for all the detected components along the same line of sight, including the ionized gas (0.2~mag), resolved stars (0.6~mag) and stellar continuum (2.0~mag).

The significant extinction difference between AT2018cow and the mini-starburst in \textit{Region-0} is of great interest. If AT2018cow was a member of the mini-starburst, the difference is possibly due to the very fractal spatial distribution of interstellar dust, such that AT2018cow happens to be obscured by no dust along the line of sight. This is, however, not supported by the very small extinction spread (0.05$^{+0.09}_{-0.02}$~mag) derived for the resolved sources in \textit{Region-0} (Section~\ref{resolved.sec}). Another possibility is that the progenitor may have a very powerful wind and have cleared any dust in the foreground before explosion. This requires the previous dust to be within striking distances at which the wind was strong enough to blow it away.


\subsubsection{Scenario 2: the progenitor was not a member of the mini-starburst}

The above-mentioned extinction difference is most easily explained if AT2018cow resides in the foreground of the mini-starburst in \textit{Region-0}, such that they have different line-of-sight distributions relative to dust (see the illustration in Fig.~\ref{struct.fig}). In this scenario, its progenitor is likely to arise from another star-forming event, which occurred at a different place and at a different epoch from the observed mini-starburst. Given the large gas reservoir, self-propagating star formation is naturally expected, giving rise to young stars with different ages and locations in this area\footnote{The large size of this star-forming area also corresponds to a large age spread of star formation, as expected for hierarchical star formation regulated by turbulence \citep{Efremov1998, Grasha2015, Grasha2017a, Grasha2017b, Sun2017a, Sun2017b, Sun2018, Miller2022}.} \citep[such as in the 30~Dor region;][]{Selman1999, Cignoni2015}.

Since massive stars almost always form in groups, it is reasonable to assume a number of stars to have been formed together with the progenitor of AT2018cow from the same star-forming event. This stellar population therefore has the same age as the progenitor, and because they cannot have moved too far away from their birthplace (i.e. in the foreground of the mini-starburst of \textit{Region-0}) during the short lifetime of a massive star, they should also have a similarly low internal extinction as AT2018cow.

Such a population is not detected on the HST images (Figs.~\ref{image.fig} and \ref{struct.fig}), and we try to constrain its age based on the detection limits. Figure~\ref{for.fig} shows the CMDs of stellar populations of log($t$/yr) = 6.9, 7.0 and 7.1, respectively, simulated with the \textsc{parsec} \citep[v1.2S;][]{parsec.ref} stellar isochrones, the \citet{imf.ref} IMF, a 50\% (non-interacting) binary fraction and a flat distribution of primary-to-secondary mass ratio; each population has an initial mass of 10$^5$~$M_\odot$ and is reddened with the Galactic extinction without any internal extinction (consistent with AT2018cow). It is clear that the population would be significantly detected if its age were younger than log($t$/yr) = 7.0 (i.e. 10~Myr). Therefore, we constrain the age of AT2018cow's progenitor to be $\gtrsim$10~Myr, corresponding to an initial stellar mass of $M_{\rm ini}$ $\lesssim$ 20~$M_\odot$ if the progenitor followed single-star evolution or if the binary interaction have not affected its lifetime significantly.

We cannot rule out the possibility that AT2018cow's progenitor was born in a low-level star-forming event, and is not associated with any major stellar population. Future observations, with deeper detection limits, will help to test these scenarios, in particular if a separate foreground population can be identified.

\subsection{The late-time brightness of AT2018cow}

\begin{figure}
\centering
\includegraphics[width=1\linewidth, angle=0]{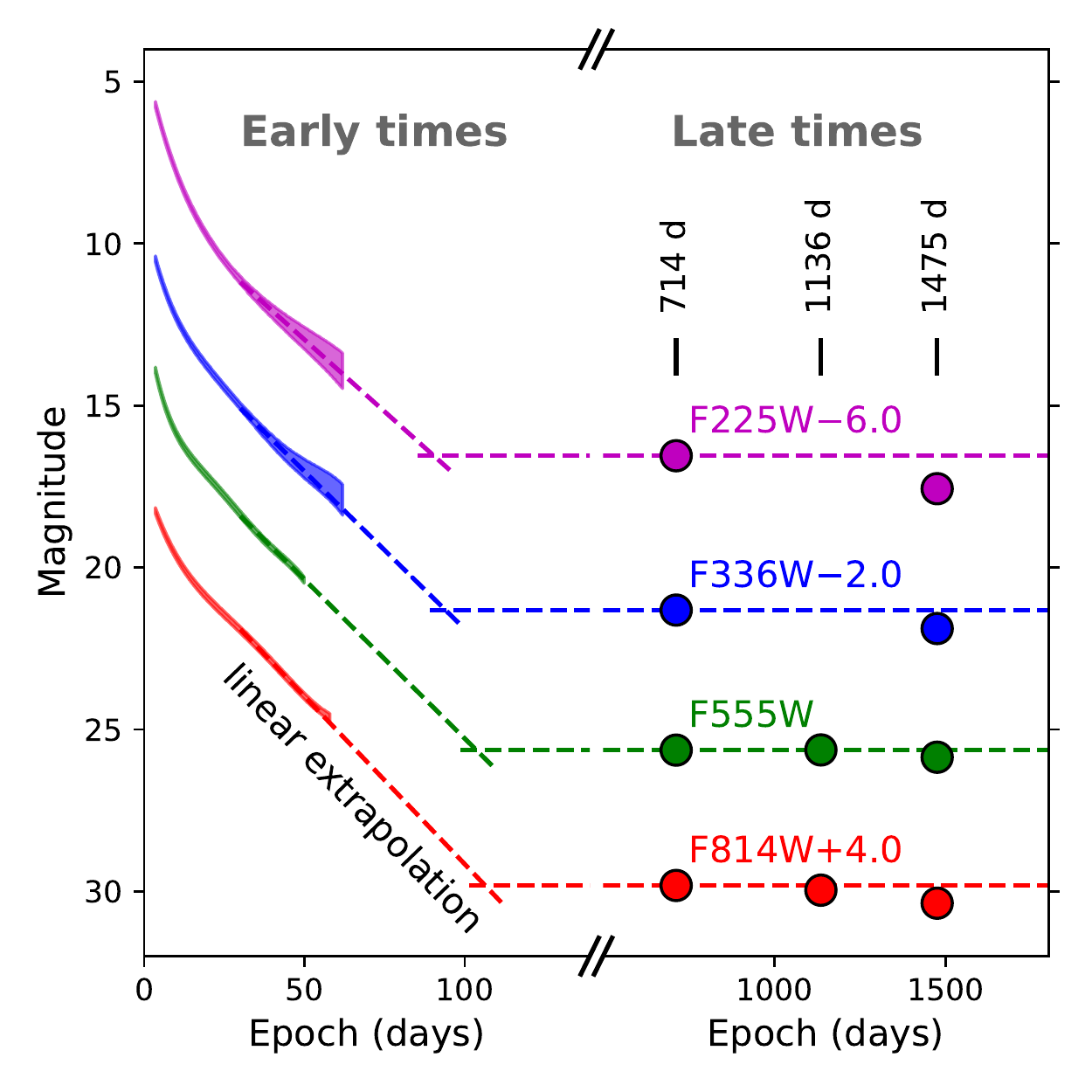}
\caption{Late-time brightness of AT2018cow (filled circles) with error bars smaller than the symbol size. The solid lines are AT 2018cow’s early-time light curves reported by \citet{Perley2019} (with line thickness showing the photometric uncertainties), which have been converted into the HST filters and the Vega magnitude system (see \citealt{Sun2022b}). The dashed lines correspond to linear extrapolations of the light curve tails or the late-time brightness at $t$ = 714~d.}
\label{curve.fig}
\end{figure}

\begin{figure}
\centering
\includegraphics[width=1\linewidth, angle=0]{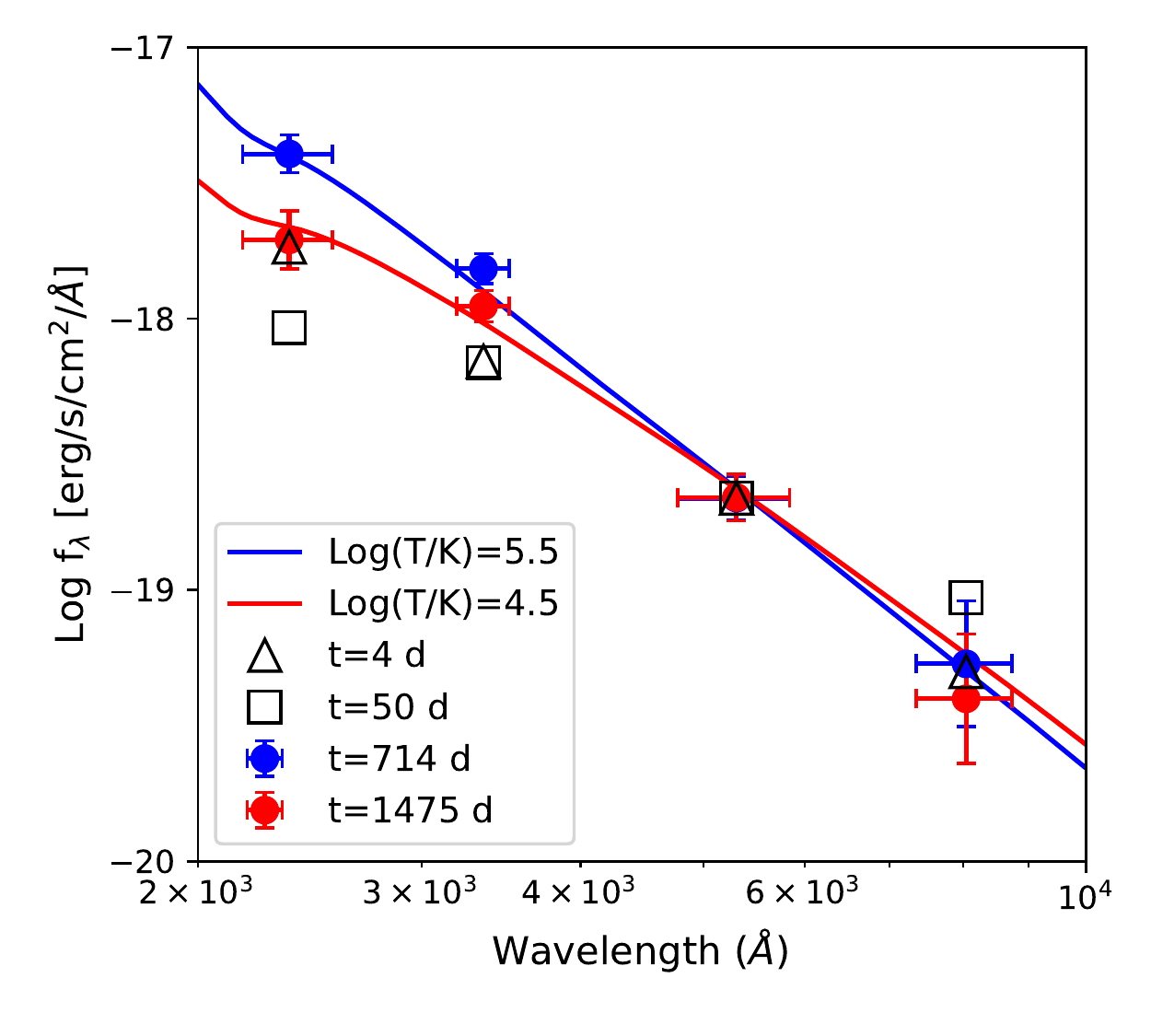}
\caption{AT2018cow’s early-time SEDs at $t$ = 4~d (open triangles) and 50~d (open squares) and late-time SEDs at $t$ = 714~d (blue filled circles) and $t$ = 1475~d (red data points), all normalized to the F555W brightness at $t$ = 714~d. The horizontal error bars correspond to the root-mean-square widths of the HST filters and the vertical error bars reflect the 5$\sigma$ photometric uncertainties. For comparison we show blackbody spectra of log($T$/K) = 5.5 (blue line) and 4.5 (red line), all reddened with AT2018cow’s Galactic reddening and normalized to the F555W band.}
\label{sed.fig}
\end{figure}

\begin{table}
\caption{Late-time brightness of AT2018cow.}
\begin{tabular}{ccccc}
\hline
\hline
Epoch & F225W & F336W & F555W & F814W \\
\hline
714 & 22.55 (0.06) & 23.32 (0.05) & 25.64 (0.07) & 25.82 (0.19) \\
1136 & -- & -- & 25.63 (0.08) & 25.96 (0.24) \\
1475 & 23.57 (0.09) & 23.89 (0.05) & 25.86 (0.07) & 26.37 (0.20) \\
\hline
\end{tabular} \\
\label{source.tab}
\end{table}

\citet{Sun2022b} reported a UV-bright late-time source at the position of AT2018cow on HST images taken at $t$ = 714~d and 1136~d, or 2--3~yr after explosion (i.e. the same data as used in this work; this source can also be seen in Fig.~\ref{image.fig}f). Its brightness is much higher than the simple extrapolation of the early-time light curves. The SED is consistent with the Rayleigh-Jeans tail of a very hot blackbody spectrum with temperature log($T$/K) $>$ 4.7 and luminosity log($L$/$L_\odot$) $>$ 7.0 (see also Fig.~\ref{sed.fig}), and its optical F555W/F814W brightness is surprisingly stable within photometric uncertainties between the two epochs. H$\alpha$ emission is also marginally detected. \citet{Sun2022b} explored the nature of this late-time source, considering the scenarios of an unrelated object in chance alignment, a companion star, the survived progenitor, the host star cluster, a light echo, or due to the late-time emission from AT2018cow itself generated by, e.g., ejecta-CSM interaction, magnetar spin-down, or accretion on to the black hole. They found that no simple models can explain all of the observed features.

The environmental analysis in this work suggests that the late-time source is not due to a stellar object. Although the stable F555W/F814W brightness is most easily explained with a star/star cluster, the source is far brighter at UV wavelengths than any other sources in \textit{Region-0} (see Fig.~\ref{cmd.fig}), and its close-to-zero internal extinction is inconsistent with the higher extinctions derived for the stars. Therefore, we can rule out this source being an object from the environment, at least at the UV wavelengths. Furthermore, \citet{Sun2022b} showed that, if the source were a cluster (or a single star), it must be very young (and very massive) in order to account for the observed UV brightness\footnote{However, even the hottest stars and youngest star clusters are still very difficult to match the very blue SED of the late-time source; see Fig.~3 of \citet{Sun2022b}.}. This requires it to be even younger than the observed mini-starburst in \textit{Region-0}, which seems to be very unlikely. In addition, if the late-time source were a close companion star of AT2018cow, the energy injection from ejecta-companion interaction may make it much more luminous than its pre-interaction state \citep{Hirai2018, Ogata2021}; however, the inflated companion will have a very low effective temperature, which is inconsistent with the observed very hot SED.

\citet{Sun2022b} disfavoured the late-time source as arising from a dust-scattered light echo, which is difficult to explain (1) the very blue SED (which is even bluer than AT2018cow at early times), (2) the very stable F555W/F814W magnitudes, and (3) the marginally detected H$\alpha$ emission (which is not seen in AT2018cow's featureless spectrum around peak brightness; \citealt{Kuin2019, Perley2019, Xiang2021}). Therefore, it seems that the brightness observed at 2--3~yr after explosion is most likely to originate from AT2018cow itself, which is somewhat surprising for a very "fast" transient that rises to peak within a few days and declines dramatically in months.

During the writing of this paper, AT2018cow was observed by HST at a new epoch of $t$ = 1475~d (29 Jun 2022) in the F225W, F336W, F555W and F814W filters (Program ID: 16925: PI: Chen Y.). At this epoch, the late-time source of AT2018cow was still bright and we performed its photometry with the \textsc{dolphot} package. The measured magnitudes are listed in Table~\ref{source.tab}, along with those at $t$ = 714~d and 1136~d (taken directly from \citealt{Sun2022b}), and displayed in Fig~\ref{curve.fig} in comparison with the early light curves. It is clear that the late-time brightness of AT2018cow has declined from $t$ = 714~d to $t$ = 1475~d, in particular in the UV filters with differences of $\Delta m_{\rm F225W}$ = 1.02~mag and $\Delta m_{\rm F336W}$ = 0.57~mag, much larger than the photometric uncertainties. The SEDs of the late-time source (Fig.~\ref{sed.fig}) has also changed, from log($T$/K) $>$ 4.7 and log($L$/$L_\odot$) $>$ 7.0 at $t$ = 714~d \citep{Sun2022b} to log($T$/K) $\sim$ 4.5 and log($L$/$L_\odot$) $\sim$ 6.4 at $t$ = 1475~d. These results support our conclusion that the late-time brightness is powered by AT2018cow itself due to some time-varying mechanisms. For example, \citet{Metzger2022} proposed that this late-time brightness could arise from the thermal emission from the accretion disk around a black hole, whose effective temperature and luminosity are expected to slowly decline with time.

\section{Summary and conclusions}

In this paper we carry out a comprehensive environmental analysis of the FBOT AT2018cow using data observed by ALMA, VLT/MUSE and HST/WFC3. The combined dataset allows us to probe the molecular gas, ionized gas and young stellar populations in the environment and investigate their relationship to AT2018cow.

The CO~(1--0) line observed by ALMA reveals a prominent concentration of molecular gas in the vicinity of AT2018cow. It has a very large mass of 6 ($\pm$ 1) $\times$~10$^6$~$M_\odot$ and provides enough material for active star formation. The molecular gas is associated with two giant star-forming complexes (\textit{Region-0} and \textit{Region-1}) rich in young massive stars/star clusters that have photoionized the surrounding gas.

By analysing the nebular emission lines and stellar continuum from the VLT/MUSE datacube and the point sources spatially resolved on the HST/WFC3 images, we find \textit{Region-0} (or \textit{Region-1}) has a very young age of 4 $\pm$ 1~Myr (or $\lesssim$2.5~Myr), a total stellar mass of 3 $\times$ 10$^6$~$M_\odot$ (same for both regions), and an H$\alpha$ luminosity of 3.4 ($\pm$ 0.5) $\times$~10$^{38}$~erg~s$^{-1}$ [or 1.64 ($\pm$ 0.05) $\times$~10$^{39}$~erg~s$^{-1}$]. All these parameters support both regions being mini-starbursts that resemble 30~Dor in the LMC.

The different internal extinctions measured for the components possibly suggest their different locations relative to dust along the line of sight. In this scenario, from the background to the foreground are the molecular gas, the majority of the newly formed stars that dominate the stellar continuum , the few most luminous and least extinct stars that are spatially resolved as point sources, followed by the ionized gas that has broken out from the molecular gas and is possibly flowing toward the observer.

AT2018cow is spatially coincident with one of the mini-starbursts, \textit{Region-0}. If its progenitor was a member of it and had the same age (4 $\pm$~1~Myr), we derived a large initial mass of $M_{\rm ini}$ = 66$^{+47}_{-22}$~$M_\odot$ if it followed single-star evolution or if binary interaction has not affected its lifetime. However, the close-to-zero internal extinction of AT2018cow is much smaller than that of the mini-starburst, and it seems more likely that its progenitor resides in the foreground and was born in a different star-forming event. The non-detection of the associated stellar population constrains its age to be $\gtrsim$10~Myr, corresponding to an initial mass of $M_{\rm ini}$ $\lesssim$ 20~$M_\odot$ assuming single-star evolution.


Finally, our environmental analysis shows that the late-time brightness of AT2018cow reported by \citet{Sun2022b} at $\geq$2~yr after explosion is unlikely to be a stellar source either from its environment or directly associated with AT2018cow (e.g. companion star, host star cluster, the survived progenitor, etc.). The brightness is most likely to arise from AT2018cow itself due to some powering mechanism working at very late times. This is also supported by the new epoch of HST observations showing that the late-time brightness has slightly declined from $t$ = 714~d to 1475~d.

\section*{Acknowledgements}

We are grateful to Prof. Paul A. Crowther for the helpful discussion during this work. Research of N-CS and JRM is funded by the Science and Technology Facilities Council through grant ST/V000853/1, and IAJ is funded by the Sheffield Undergraduate Research Scheme. This paper is based on observations made with the NASA/ESA Hubble Space Telescope (Programs 15600, 15974 and 16179) and the Very Large Telescope at ESO (Program~0103.D-0440). This paper also makes use of the following ALMA data: ADS/JAO.ALMA$\#$ 2017.A.00045.T. ALMA is a partnership of ESO (representing its member states), NSF (USA) and NINS (Japan), together with NRC (Canada), MOST and ASIAA (Taiwan), and KASI (Republic of Korea), in co-operation with the Republic of Chile. The Joint ALMA Observatory is operated by ESO, AUI/NRAO and NAOJ. The National Radio Astronomy Observatory is a facility of the National Science Foundation operated under cooperative agreement by Associated Universities, Inc.

\section*{Data availability}
All data are publicly available from the ALMA science archive (\url{https://almascience.nrao.edu/}), the ESO data archive (\url{http://archive.eso.org/}), and the Mikulski Archive for Space Telescopes (\url{https://archive.stsci.edu/}).

\bsp	
\label{lastpage}
\end{document}